\documentclass[10pt,conference]{IEEEtran}
\IEEEoverridecommandlockouts
\usepackage{cite}
\usepackage{amsmath,amssymb,amsfonts}
\usepackage{algorithmic}
\usepackage[ruled,vlined]{algorithm2e}
\usepackage{graphicx}
\usepackage{textcomp}
\usepackage{xcolor}
\usepackage{url}
\usepackage{footnote}
\usepackage[perpage]{footmisc}
\usepackage{balance}
\usepackage{multirow}
\usepackage{listings}

\def\BibTeX{{\rm B\kern-.05em{\sc i\kern-.025em b}\kern-.08em
    T\kern-.1667em\lower.7ex\hbox{E}\kern-.125emX}}
    
\begin{document}

\newcommand{\eg}{\textit{e.g., }}
\newcommand{\ie}{\textit{i.e., }}
\newcommand{\al}{\textit{et al. }}

\newcommand{\Foutse}[1]{\textcolor{red}{{\it[Foutse:#1]}}}
\newcommand{\Amin}[1]{\textcolor{blue}{{\it[Amin:#1]}}}
\newcommand{\PO}[1]{\textcolor{orange}{{\it[PO:#1]}}}
\newcommand{\Rached}[1]{\textcolor{brown}{{\it[Rached:#1]}}}

\title{Quality issues in Machine Learning Software Systems\\
%\thanks{Identify applicable funding agency here. If none, delete this.}
}

\author{\IEEEauthorblockN{Pierre-Olivier Côté, Amin Nikanjam, Rached Bouchoucha, Foutse Khomh}
\IEEEauthorblockA{\textit{SWAT Lab., Polytechnique Montréal, Québec, Canada} \\
\{pierre-olivier.cote,amin.nikanjam,rached.bouchoucha,foutse.khomh\}@polymtl.ca}
%}
}

%\and
%\IEEEauthorblockN{6\textsuperscript{th} Given Name Surname}
%\IEEEauthorblockA{\textit{dept. name of organization (of Aff.)} \\
%\textit{name of organization (of Aff.)}\\
%City, Country \\
%email address or ORCID}
%}

\maketitle

\begin{abstract}
Context: An increasing demand is observed in various domains to employ Machine Learning (ML) for solving complex problems. ML models are implemented as software components and deployed in Machine Learning Software Systems (MLSSs). Problem: There is a strong need for ensuring the serving quality of MLSSs. False or poor decisions of such systems can lead to malfunction of other systems, significant financial losses, or even threat to human life. The quality assurance of MLSSs is considered as a challenging task and currently is a hot research topic. Moreover, it is important to cover all various aspects of the quality in MLSSs. Objective: This paper aims to investigate the characteristics of real quality issues in MLSSs from the viewpoint of practitioners. This empirical study aims to identify a catalog of bad-practices related to poor quality in MLSSs. Method: We plan to conduct a set of interviews with practitioners/experts, believing that interviews are the best method to retrieve their experience and practices when dealing with quality issues. We expect that the catalog of issues developed at this step will also help us later to identify the severity, root causes, and possible remedy for quality issues of MLSSs, allowing us to develop efficient quality assurance tools for ML models and MLSSs.
\end{abstract}

\begin{IEEEkeywords}
Machine Learning based Software Systems, Quality Assurance, Quality issues, Interview.
\end{IEEEkeywords}

\section{Introduction}
Nowadays, Machine Learning Software Systems (MLSSs) have become a part of our daily life (e.g., recommendation systems, speech recognition, face detection). An increasing demand is observed in various companies to employ Machine Learning (ML) for solving problems in their business. Typically, a MLSS receives data as input and employs ML models to make intelligent decisions automatically based on learned patterns, associations and knowledge from data \cite{marijan2019challenges}. Therefore, ML models are implemented as software components integrated to other subsystems in MLSSs, and like other software systems, quality assurance is necessary. According to the growing importance of MLSSs in today’s world, there is a strong need for ensuring their serving quality. False or poor decisions of such systems can lead to malfunction of other systems, significant financial losses or even threat to human life \cite{foidl2019risk}. %Figure 1 illustrates the life cycle of a ML model in a MLSS.
 
The quality assessment of MLSSs is regarded as a challenging task \cite{braiek2020testing} and currently is a hot research topic \cite{khomh2018software,breck2019data}. Recently some research work on the quality of MLSSs is suggested to cover all different aspects of the quality \cite{studer2021towards,breck2017ml}, i.e., not only their prediction accuracy. In this paper, we, with our industrial partner, plan to investigate the characteristics of real-world quality issues in MLSSs from the viewpoint of practitioners identifying a list of bad-practices related to poor systems/models quality. This is a requirement of comprehensive quality assessment of MLSSs, as Zhang \al already acknowledged the lack of such empirical study and asserted that conducting empirical studies on the prevalence of poor models among deployed ML models should be interesting \cite{zhang2020machine}. This study will cover all relevant quality factors like performance (accuracy), robustness, explainability, scalability, hardware demand and model complexity. We plan to conduct a set of interviews with practitioners/experts, believing that interviews are the best method to retrieve their experience and practices when dealing with quality issues. We expect that the catalog of issues developed at this step will also help us later to identify the severity, root causes and possible remedy for quality issues of MLSSs, allowing us to develop efficient quality assurance tools for ML models and MLSSs. We present in the following the proposed methodology to achieve our objectives.

%We aim through this paper to investigate if our catalog of smells is prevalent. We also aim to investigate its severity and impact on software quality attributes. To achieve our objectives, we analysed nine open source projects to detect occurrences of multilanguage design smells. We plan to conduct a survey, we believe that a survey is the best method to retrieve developers’ experience and practices when dealing with multi-language systems [5]. We present in the following the proposed methodology to achieve our objectives.
% While the last decade has seen great improvement in model architectures \cite{DBLP:journals/corr/VaswaniSPUJGKP17, krizhevsky2012imagenet}, researchers and practitioners are slowly moving toward a data-centric approach to machine learning \PO{find citation}. In fact, 80-90\% of the machine learning process is spent on data preparation \cite{whang2021data}.
%\PO{It could be worthwhile to situate our work in the data-centric AI trend. There is hype on this subject since Andrew Ng viewed it as a very promising branch of AI.}
\section{Related Work}
%\Foutse{ I think you will need to discuss related Works to explain why current literature is incomplete and why your qualitative study is needed to complement current knowledge about quality issues in ML data}

In 2015, Sculley \al shared through a seminal work, the challenges that Google faced while building, deploying, and maintaining ML models \cite{NIPS2015_86df7dcf}. Following this work, many other studies tried to characterize quality issues in MLSSs \cite{washizaki2019studying, alahdab2019empirical, bogner2021characterizing, amershi2019software, sculley2014machine, van2021prevalence, tang2021empirical, nahar2022collaboration }. Van Oort \al mined open-source GitHub repositories to aggregate a list of the most common maintenance-related modifications in Deep Learning (DL) projects \cite{bogner2021characterizing}. From this list, they extracted 5 code smells in DL systems. Furthermore, they measured how prevalent and problematic the code smells are from the point of view of practitioners using a survey. Similarly, Dilhara \al \cite{dilhara2021understanding} mined 26 open-source MLSSs and identified 14 refactoring and 7 new technical debt (TD) categories specific to ML. Instead of finding new quality issues, Alahdab \al shared with the scientific community how TD types appear in the early phases of an industrial DL project \cite{alahdab2019empirical}. Facing the increasing number of papers regarding technical debt and code smells in MLSSs, Bogner \al attempted to aggregate the knowledge on these topics by performing a systematic mapping study that presented 4 new types of technical debt, 72 antipatterns along with 46 solutions \cite{bogner2021characterizing}. A similar work has been done by Washizaki \al but by consulting grey literature as well \cite{washizaki2019studying}. Other works indirectly contributed toward increasing the knowledge on quality issues in MLSSs, such as the study of Nahar \al \cite{nahar2022collaboration}, which looked at collaboration challenges while building ML systems. In reaction to the growing concern of quality issues in MLSSs, researchers developed tools such as the DataLinter \cite{hynes2017data} or the data validation component of TFX \cite{breck2019data} to automate quality assurance processes. Another group of researchers adopted a different approach and instead shared a checklist of tests to assess the production readiness of MLSSs \cite{breck2017ml}. In a similar fashion, a process model is proposed by Studer \al for the development of ML applications with a quality assurance methodology \cite{studer2021towards}. While there is an increasingly growing number of studies on quality issues in MLSSs, a significant portion of them is produced by large software enterprises such as Google or Meta, which limits the generalization of the findings \cite{bogner2021characterizing}. We believe that there is a need for a study presenting the quality issues of MLSSs that are encountered by practitioners from different backgrounds and company sizes.

% There exists many approaches to solve these data quality issues. One is to prevent them from entering the ML system. \cite{breck2019data} is an example of a data validation tool. An other approach, named data cleaning, tries to fix faulty data instead. While the field is studied since a long time in the DB community \cite{hellerstein2008quantitative, 10.1145/2882903.2912574}, recent work \cite{li2019cleanml} showed that using these techniques can degrade the performances of the end model instead of improving them. In order to avoid this situation, studies started using the end model performance to guide the cleaning process \cite{krishnan2017boostclean, krishnan2016activeclean}. Other ones use ML to predict repair on faulty instances \cite{rekatsinas2017holoclean}. The field of data cleaning and machine learning is emerging and there is a need to document the existing approaches. Previous work either discussed data cleaning for ML in the context of a larger subject \cite{roh2019survey} or only briefly described the recent progress \cite{neutatz2021cleaning}. There is a need to have a study thoroughly reviewing the existing approaches of data cleaning in the context of Machine Learning.

\section{Study Design}
\subsection{Objectives and Research Questions}
%\PO{TODO: find RQs for quality issues in general}
The goal of this study is to provide a detailed analysis of quality issues (including data and model) in MLSSs. We believe that interviewing people who really experienced these issues is an effective way to gain that knowledge. Thus, we will proceed this way and we define the following Research Questions (RQ):
\begin{description}
    \item[RQ1:] \textbf{What are the quality issues encountered by practitioners when building MLSSs and which ones are the most prevalent?} For future works to solve quality issues, they must know the issues that exist. While there is some literature covering some quality issues \cite{NIPS2015_86df7dcf, nahar2022collaboration,studer2021towards}, we believe that a lot is still unknown. In this study, we aim to share with the research community the quality challenges encountered by practitioners when building ML systems. This includes issues related to data, model, and other components in MLSSs.
    
    \item[RQ2:] \textbf{What are the root causes, symptoms, and consequences of quality issues?} To understand quality issues in ML systems, it is not sufficient to know they exist, but also their root causes, symptoms and potential consequences. We hope that the answer to that question will guide future work towards solving the most pressing issues.
    
    \item[RQ3:] \textbf{How are the quality issues currently handled by the practitioners?} Once quality issues have been detected (e.g., in data or model), we expect practitioners to have put in place mechanisms to mitigate or at least attenuate their consequences. We are interested in understanding the current mitigation approaches implemented in the industry. %they adopted to do so. 
    
    %\item[RQ4:] \textbf{From which data quality dimension most data quality issues come from?} To define data quality issues, one must define what is data quality. This concept varies from one application to another and is often unclear \cite{neutatz2021cleaning}. For example, a dataset where string attributes inconsistently begin with a capital letter can be a problem for a ML model, but it is not for a human. Thus, in one situation, the dataset would have quality issues, while in the second one, not. Previous research did an effort in the definition of dataset quality dimensions for ML models \cite{cappi2021dataset}. In this research project, we will clarify which quality dimensions are the most troublesome for practitioners when assessing the quality of their ML system.
    
    \item[RQ4:] \textbf{In the case of data, which data types and collection processes are the most challenging in terms of data quality?} The training data ingested by ML models comes in many forms. Face recognition systems use images, while stock prediction applications can use numbers/amounts (like time-series data) or even text. Each data type comes with its own data quality challenges when training ML models. By answering this RQ, future research will be guided towards the most pressing data quality challenges faced by practitioners. Authors in \cite{whang2021data} mention that data collection processes, the process of gathering data, may affect the quality of data. For example, one could choose to train her/his model on public datasets, or manually acquire more data with data collectors. Each of these processes have different challenges which may lead to different data quality issues. In this study, we want to identify %put in broad daylight 
    the data collection processes that are most prone to %where practitioners encounter the most 
    data quality issues.
    
    %\item[RQ6:] \textbf{Which data types are the most challenging in terms of data quality?} The training data ingested by ML models comes in many forms. Face recognition systems use images, while stock prediction applications can use numbers/amounts (like time-series data) or even text. Each data type comes with its own data quality challenges when training ML models. In this study, we will present the data quality challenges faced by practitioners for each data type. By answering this RQ, future research will be guided towards the most pressing data quality challenges faced by practitioners.
    
    \item[RQ5:] \textbf{What are the challenges of data quality assurance during model evolution? %are quality issues occurring as time goes by?
    } Many MLSSs encounter rapidly changing/non-stationary data, adversarial input, or difference in data distribution (concept drift, e.g., content recommendation systems and financial ML applications). Hence, the quality of the deployed model may be decreasing over time and consequently affects the performance of the whole system. Therefore, the robustness and the accuracy of the model's predictions must be assessed frequently in production. Actively monitoring the quality of the deployed model in production is crucial to detect performance degradation and model staleness.
    
\end{description}

% Furthermore, we are interested in the existing approaches that can be used to solve these issues. More specifically, we want to summarize the body of literature on the topic of data cleaning in the advent of machine learning. Hence, we define the following research questions.
% \\\\
% \textit{RQ2: What are the existing approaches to repair data quality issues for machine learning models?}
% \\\\
% By answering these questions, we think we will provide researchers and practitioners knowledge necessary to improve the quality of machine learning models by enhancing the quality of data. \PO{Maybe we should not say "to repair data quality issues" in the RQ because we are not fixing every possible data quality issue, only cleaning some type of issue (data cleaning can not fix data traceability issues for example)}

% \item \textbf{How are the data quality issues currently addressed by the practitioners?\Amin{and available tools?}}
% \item \begin{enumerate}\Amin{I think this should be merge to the previous RQ!}
%     \item How are these quality issues addressed in the literature?\Amin{looks the same as the previous one!} 
%     \item Which techniques can be used to fix the aforementioned data quality issues?
%     \item How can data quality issues can be fixed in the context of ML?\Amin{our context is already ML!}
% \end{enumerate}

\subsection{Industrial Partner}
MoovAI\footnote{\url{https://moov.ai/en/}} is a Montreal-based company in Canada, which is active in developing AI/ML quality assurance solutions to address practical needs in the various businesses. MoovAI’s experts guide their customers (i.e., companies) to take advantage of these cutting-edge technologies, regardless of their level of maturity in data science. Nowadays, various sectors in industry are and will be developing ML models in their systems, e.g. energy sector, financial sector, supply chain recommendations, medical diagnosis and treatment. As ML-based technologies become more widespread, the quality demands of ML become more important. Poor quality models are a potential barrier in exploiting ML in real-world applications and currently more companies request for qualified ML-systems \cite{CD4MLSato}. A study showed that 87\% of ML proof of concepts have never come to production \cite{azimi2020root}, usually due to lack of enough quality.

Currently MoovAI is using a model validation tool to validate ML models prior to deployment in the real world \cite{Blais2020}. This tool includes preliminary validation methods to assess the overall quality of a ML model. The tool evaluates accuracy, stability, biases and sensitivity of ML models. Now, MoovAI aims to push their tool forward to develop advanced methods for ensuring the quality of not only the ML models but also ML-based systems (i.e., software systems containing ML components) and to develop a stand-alone model validation platform. Experts at MoovAI now are looking for a comprehensive quality evaluation tool to assess and monitor the quality of ML models during their whole life cycle; from data collection, to development, deployment, and maintenance. 

\subsection{Participants}  \label{section:participants}
We plan to interview at least 50 participants, which is more than previous similar studies \cite{humbatova2020taxonomy, serban2021empirical}. The interviews will occur in two rounds. In the first round of the study, we will interview the employees of our industrial partner (i.e., MoovAI). They are Data Scientists, Machine Learning Engineers, Data Engineers and Project Managers of ML projects. They have worked on many different ML projects for different clients. Thus, we think that they are able to provide a global view of quality issues encountered in the industry. We expect to be able to conduct at least 25 interviews with MoovAI's practitioners. In the second phase of the study, we plan to interview practitioners from other companies. We will adopt 4 strategies to recruit participants.
\begin{itemize}
    \item \textit{Personal contacts}: We will start recruiting people for interviews through our personal contacts, since it usually has a higher response rate than cold emails. We will contact industrial partners in our network. For example, we plan to reach out to companies who are partners in Software Engineering for Artificial Intelligence\footnote{\url{https://se4ai.org/}}, a training program co-created by our lab which includes companies such as IBM, Ericsson, and Cisco. We will also use LinkedIn\footnote{\url{https://www.linkedin.com/}} to find qualified experts that may have relevant expertise for this project. Using the results from a similar study \cite{humbatova2020taxonomy}, we expect at least 5 positive responses.
    
    \item \textit{Q\&A websites}: Questions and Answers websites are platforms on which a lot of knowledge is shared. Following similar previous work \cite{humbatova2020taxonomy}, we will search for practitioners with meaningful experience on ML willing to be interviewed on quality issues in ML systems. We chose to search on Stack Overflow\footnote{https://stackoverflow.com/} and Data Science Stack Exchange\footnote{https://datascience.stackexchange.com/}, because both these websites are significantly used by the ML community for question answering. We will reach out to the top askers and answerers\footnote{The words \textit{askers} and \textit{answerers} are part of the terminology used by the Q\&A websites to describe people asking and answering questions} since they have shown significant involvement in the ML community and most likely have expertise. In order to find them, we will search on two platforms: 1) Data Science Stack Exchange, on which we will simply look for the top answerers and askers of the Q\&A platform on any topic, and 2) Stack Overflow, for which a different strategy must be adopted since the website also holds questions regarding Software Engineering in general. Thus, we will search for the top askers and answerers for topics related to the subject of quality issues in ML systems, using tags. Similar to \cite{humbatova2020taxonomy}, we selected the tags \textit{data-cleaning, dataset, machine-learning} and \textit{artificial-intelligence}. We will search for practitioners on Data Science Stack Exchange and on Stack Overflow using the 4 tags we have described (1 for Data Science Stack Exchange + 4 tags on Stack Overflow, 5 in total). Top users are ranked in two categories: \textit{'Last 30 Days'} and \textit{'All Time'}; we will pick the first 10 users from both. In total, we will send 100 emails. Using the results from a similar study \cite{humbatova2020taxonomy}, we expect to meet 10 interviewees.

    \item \textit{Social networks}: Social networks are platforms in which users may engage conversations on a wide range of topics. Some of them host discussions on ML and Artificial Intelligence in general. We believe users with important ML expertise can be found on these websites. We are planning to post an invitation for interviews on the deep learning and ML communities of Reddit\footnote{\url{https://www.reddit.com/}} similar to our previous works \cite{nikanjam2021design}. We expect at least 5 positive responses.
    \item \textit{Freelance platforms}: Following previous work \cite{humbatova2020taxonomy}, we plan to use freelance platforms to find practitioners with ML expertise if all the other techniques have been used and we have not reached theoretical saturation yet \cite{stol2016grounded}. We will follow the methodology of similar studies like \cite{humbatova2020taxonomy} to select the candidates for interviews.
\end{itemize}

\subsection{Interview Process}

Since the subject of quality issues is not mature and has still room for research work, we will follow a research procedure suited for exploratory work, Straussian Grounded Theory \cite{strauss1998basics}. It is a research method in which data collection and data analysis are executed in an iterative manner until a new theory emerges from the data. As opposed to many deductive approaches where a theory is first conceptualized then tested through experiments, Grounded Theory goes the opposite way by inductively generating a theory from the data. The knowledge gathered from data using open and axial coding should guide the sampling process, a concept named theoretical sampling. Data collection stops when theoretical saturation is met: when the understanding of the subject is complete and new data does not invalidate the emerging theory. Grounded Theory is often used when little is known about a phenomenon, because of its flexibility and its aptitude for the discovery of unknown concepts. It has been used in similar studies, such as \cite{nahar2022collaboration}. Because we are doing exploratory work, the interviews have to be structured in a way that allows the interviewee to share knowledge we might not be aware of. For this reason, we will be conducting semi-structured interviews similar to \cite{humbatova2020taxonomy}. We devised an interview guide that to help the interviewer cover every relevant topic. It is composed mostly of open-ended questions, to allow the interviewee direct us towards interesting information. It is the responsibility of the interviewer to ask follow-up questions when the respondent touches upon a subject relevant to the study. To avoid the interviewer being overwhelmed with his tasks and having difficulty asking the right follow up questions, each interview will be conducted by two persons. This follows the recommendation of previous works which show that interviewees share more information when two interviewers are present rather than one \cite{1509301}. The second interviewer will be tasked with helping the primary one to ask follow-up questions and to transcribe the interview with the support of the tool Descript\footnote{\url{https://www.descript.com/transcription}}. This tool is an automated speech recognition tool that can transform the audio stream of an online meeting into text. The role of the second interviewer will simply be to correct the mistakes of the transcription of the tool.

The interview guide will be subject to change as our knowledge on the topic grows, since we are following Grounded Theory. During its writing, we will consider quality issues elicited by other studies \cite{NIPS2015_86df7dcf, serban2021empirical, bogner2021characterizing, washizaki2019studying, breck2017ml}. We also took care of asking questions that have potential to cover different quality dimensions as defined by \cite{cappi2021dataset}. We will complete our interview guide with questions drawn from a similar study \cite{humbatova2020taxonomy}. In order to assess the quality of our interview guide, we will conduct a pilot of our study. We will purposefully select 5 participants with diverse experience (i.e. Data scientists, Data/ML Engineers and AI project managers). Doing so, we will be able to verify that our questions are unambiguous, precise and able to answer our research questions effectively. A few days prior to the interview, we will share a summary of the content interview with the participants so they can be familiar with the objectives of the study. 
 
Interviews will take place in English or French, depending on the preference of the interviewee. We will start by giving a brief overview of the project followed by a quick round of introductions. Then we will ask participants for some background information regarding his experience in ML. The interview will officially start with a general question: "\textit{What are the main quality issues you have encountered with your data/model so far}"? By asking an open-ended question, we are allowing the interviewee to share experiences he is the most confident to talk about. Then we will probe the interviewee’s experience in an attempt to discover quality issues. We will cover each phase of a ML workflow as described by \cite{amershi2019software}. in order to exhaustively search for situations where quality issues might occur. As a difference, we do not include the model requirement phase in our study and chose to merge data cleaning with feature engineering and data labeling with data collection for conciseness.
\begin{itemize}
    \item \textit{Data collection}: We ask for experiences with different data collection processes: data collectors, automatically generated data, public datasets, external services (e.g. a weather API), or predictions of another model (effectively creating cascading models). For each one of them, we search for issues the interviewee may have experienced collecting the data along with solutions they put in place.
    \item \textit{Data preparation}: Notably, we ask about pain points when preparing data for ML and for tools to automate the process. We consider any challenge related to the cleaning and transformation of the data.
    \item \textit{Model evaluation}: We probe for potential problems that the interviewee experienced when evaluating the quality of its model. For example, we ask if the respondent ever faced a situation where the model performed poorly on some group of people, potentially leading to fairness issues.
    \item \textit{Model deployment}: We gather general information about the process by which models are put into production. Then, we search for issues encountered at this step. For example, one question is: “\textit{Did you ever deploy a model that performed well locally but poorly once deployed?}”.
    \item \textit{Model maintenance}: We ask the interviewee how he ensures that the quality of its models remains the same after deployment. We specifically ask for past instances of model staleness and how it has been handled.
\end{itemize}
At the end of every section, we ask for any other issue that the interviewee may have at this step of the workflow in case we missed out on something. We will conclude the interview with the open question: "\textit{In your opinion, what is the most pressing quality issue researchers should work on in an attempt to solve the problem?}". The answer to that question might provide interesting future work directions and follows Harvard's best practices for qualitative interviews\footnote{\url{https://sociology.fas.harvard.edu/files/sociology/files/interview_strategies.pdf}}. The interview guide is available online \footnote{\url{https://github.com/poclecoqq/quality_issues_in_MLSSs}}.
 
% The second section of specific questions will be used in the interview only if the respondent actively contributes to the development of machine learning systems. It includes questions regarding model development and performance issues caused by problems in the data. For example, one question asks about past experiences with model staleness (for which the root cause of the degradation of performance is outdated training data). Questions in this section have been written down using model and data issues elicited by previous studies \cite{NIPS2015_86df7dcf} \cite{serban2021empirical}. We also designed the questionnaire for most of the dimensions of data quality \cite{cappi2021dataset} to be addressed during the interviews.  
Prior to starting the interview, we will ask to the respondent for the permission to register and share the transcription. In order to follow ethical guidelines, we validated our research project with Research Ethics Committee at Polytechnique Montréal\footnote{\url{https://www.polymtl.ca/recherche/la-recherche-polytechnique/exigences-deontologiques/travaux-de-recherche-avec-des-etres-humains}} and got their approval. We will anonymize the respondent of the interviews in the transcript in order to respect their privacy.

\subsection{Questionnaire}
After interviews have been conducted and analyzed, we expect to have a set of potential quality issues. In order to validate the quality of our findings, we will investigate their prevalence in real MLSSs. Owners of such systems will be contacted and asked to answer a questionnaire where quality issues are presented. The questionnaire will be built using Google Forms\footnote{\url{https://www.google.ca/forms/about/}}. We will reach out to owners of MLSSs by contacting MoovAI's clients. We expect to have answers from at least 20 respondents from 4 different companies. Because we are aware that they might not have a profound understanding of ML, the form will describe the symptoms of the potential quality issues including their technical description. The respondents will evaluate on a Likert scale \cite{joshi2015likert} how often this kind of problem happens in their experience. In the case of a positive response, they will be invited to share in more detail about the issues that they experienced and to describe their consequences in MLSSs.

\subsection{Analysis Plan}
For all the research questions except \textbf{RQ4}, we will use the coding techniques from Straussian Grounded Theory \cite{strauss1994grounded} to extract knowledge from data. When an interview transcript is coded for the first time, we will use open coding \cite{seaman1999qualitative} to break it into discrete parts. These codes will be reorganized and grouped into categories through a step of axial coding. This process allows the researchers to analyze its data on a higher level so to have a better understanding of it. In the final rounds of coding, the central theme around which all categories relate can be established in the final procedure of selective coding. While the current document has been written such as the data collection and analysis plan are separated, it is important to understand that these two steps will happen in an iterative manner. We expect to write down memos of preliminary categories throughout the process of data collection and analysis. They will contribute to the rise of our theory through phases of memo sorting. 
 
For \textbf{RQ4}, because the question is less open-ended, we will adopt a simpler coding strategy. The codes will be the data set quality dimensions defined in \cite{cappi2021dataset} and the data collection processes we will encounter in our interviews. We will sum the codes throughout the interviews and we will take into account duplicates: when two interviewees are on the same team and share the same problem. 
 
In order to measure the most prevalent quality issues and effectively answer \textbf{RQ1}, we will average the results (that were on a Likert scale) from the questionnaire.
 
To help us code the transcripts, we will use Delve qualitative analysis tool\footnote{\url{https://delvetool.com/}}, because the researchers are familiar with it and it is easy to use. Delve is a computer-assisted qualitative data analysis software (CADQAS) that provides simple interfaces to code and to analyze data. In order to ensure the quality of the analysis, each document will be coded by two researchers. In case of a disagreement in codes, a third researcher will play the role of moderator and will select the final code for a text segment. This process will be helpful for the construction of a shared understanding of the data.

On completion of the study, we will share with the public a replication package. This package will contain the interview guide, the anonymized transcription of the interviews and any discussion or analysis between the contributors that could help replicating the study.

\subsection{Threats to validity}
Using some of the validity threats described in \cite{feldt2010validity}, we will divide the analysis of the limitations of the study into four categories: internal validity, external validity, construct validity and conclusion limitations. The former refers to the degree of confidence that the findings are trustworthy. For example, confounding factors or a lack of scientific rigor could hinder the quality of the results. In our case, we identified four potential threats to internal validity. First, there is a risk of some quality issues being over-represented. In fact, we are conducting semi-structured interviews; some of the conversations with the interviewee will be improvised. Thus, it is possible that the findings are tainted by the preconception of the interviewer about the potential quality issues in MLSSs. For example, there could be more emphasis on explainability issues in the interview, which may lead to an over-representation of these issues in our findings. Second, it is possible that the questions asked when the interviews are in English are not exactly the same as the ones asked when the interviews are in French. While our primary interviewers have a good understanding of both languages, their choice of words might convey slightly different meanings, leading to bias. Third, the coding of the interview may be prone to researcher bias, because it partially relies on subjective interpretation. To mitigate this bias, each interview will be coded by two researchers, and inconsistencies in the codes will be resolved by a third researcher. Fourth, there is a risk that our findings for \textbf{RQ2} are inaccurate because of our data collection method. Asking practitioners about the root cause of quality issues may lead to misunderstandings or even misinformation in case they do not want to admit their mistakes. To mitigate this issue, we will cross our findings with MoovAI knowledge on similar cases.  \\
For external validity, we identified two potential ways that our sample of practitioners could be unrepresentative of the industry. First, we expect a majority of the interviewees to be practitioners from Moov.AI. Their clients are companies of medium to large size. Thus, our results may not reflect the reality of very small companies, such as startups, or larger software enterprises, such as Google, Microsoft, and the likes. Second, because the study follows Grounded Theory and we are doing theoretical sampling \cite{stol2016grounded}, it is possible that we end up focusing on some slice of the population and leaving others understudied.\\
About construct validity, limitations of our approach could come from keywords (tags) used for finding candidates over Q\&A websites and social media. For keywords, we used \textit{data-cleaning, dataset, machine-learning} and \textit{artificial-intelligence} which are sufficiently general to match a lot of users and is more likely to generate False Positive (users that we would end up ignoring) rather than False Negative.\\
Conclusion limitations can be of potential wrongly understood issues, missing issues, and the replicability of the study. We believe that our sources of information are sound and various enough to be representative, and not misleading us in our conclusions. At last, we provided a replication package\footnote{\url{https://github.com/poclecoqq/quality_issues_in_MLSSs}} to allow for the reproducibility of our results, and also for other researchers to build on our study.

% \subsection{Figures and Tables}
% \paragraph{Positioning Figures and Tables} Place figures and tables at the top and
% bottom of columns. Avoid placing them in the middle of columns. Large figures and tables may span across both columns. Figure captions should be below the figures; table heads should appear above the tables. Insert figures and tables after they are cited in the text. Use the abbreviation ``Fig.~\ref{fig}'', even at the beginning of a sentence.\cite{rensink2004groove}

% \break

% \begin{table}[htbp]
% \caption{Table Type Styles}
% \begin{center}
% \begin{tabular}{|c|c|c|c|}
% \hline
% \textbf{Table}&\multicolumn{3}{|c|}{\textbf{Table Column Head}} \\
% \cline{2-4} 
% \textbf{Head} & \textbf{\textit{Table column subhead}}& \textbf{\textit{Subhead}}& \textbf{\textit{Subhead}} \\
% \hline
% copy& More table copy$^{\mathrm{a}}$& &  \\
% \hline
% \multicolumn{4}{l}{$^{\mathrm{a}}$Sample of a Table footnote.}
% \end{tabular}
% \label{tab1}
% \end{center}
% \end{table}

% \begin{figure}[htbp]
% \centerline{\includegraphics{fig1.png}}
% \caption{Example of a figure caption.}
% \label{fig}
% \end{figure}

\section*{Acknowledgment}
This work is partly funded by the Natural Sciences and Engineering Research Council of Canada (NSERC), PROMPT, and Les Technologies MoovAI Inc. 
%We would like to thank Mr. Mohammad Mehdi Morovati who helped authors for conducting the survey.

\balance
\bibliography{references}
\bibliographystyle{ieeetr}

\end{document}